\documentclass[11pt,twoside]{article}
\usepackage{macroaaa-eng}
\usepackage{psfig}
\usepackage{graphicx}

\usepackage[T1]{fontenc} % Computer Modern (CM) fonts

\usepackage{latexsym}
\usepackage{verbatim}

\begin{document}

\vskip 1.0cm
\markboth{V. Bosch-Ramon}{The physics of non-thermal radiation in microquasars}
\pagestyle{myheadings}

\vspace*{0.5cm}
\title{The physics of non-thermal radiation in microquasars}

\author{Valent\'i Bosch-Ramon}
\affil{Max-Planck-Institut fur Kernphysik, Saupfercheckweg 1, 69117 Heidelberg, Deutschland, 
vbosch@mpi-hd.mpg.de}

\begin{abstract} 
Microquasars are binary systems that harbor a normal star and a compact object (black-hole or neutron star), and show relativistic outflows
(or jets). The matter that forms these jets is of likely stellar origin, previously expelled from the star and trapped in the potential well
of the compact object. This matter is accreted by the compact object,
forming a disk due to its angular momentum, and is eventually ejected in the form of a bipolar outflow (the
jets), which generates radio emission and could also be a very high-energy emitter. To study and understand the radiation from microquasars,
there is a set of elements that can play a major role and are to be taken into account: the photons and the expelled matter from the star in
the case of high-mass systems; the accreted matter radiation; the jet; the magnetic field carried by the jet or filling the binary system;
and the medium surrounding the microquasar at large scales ($\sim$~pc). In this lecture, we consider these elements of the microquasar 
scenario and briefly describe the
physical conditions and processes involved in the production of non-thermal radiation from radio to gamma-rays. The required
energetics, particle acceleration and transport, several radiative mechanisms, and the impact of different photon 
absorption processes, are discussed.
\end{abstract}

\section{Introduction}\label{intro}

Among the different classes of astrophysical sources, microquasars (see Mirabel \& Rodr\'iguez 1999 for a review) are specially interesting for
the study of different topics of high energy astrophysics. The fact that these objects present non-thermal emission in different wavelengths, from
radio (e.g. Rib\'o 2005) to infrared wavelengths (e.g. Mirabel et al. 1998), and from X-rays (e.g. Corbel et al. 2002) to very high-energy (VHE) gamma-rays
(Aharonian et al. 2005, 2006a; Albert et al. 2006; Albert et al. 2007), has several implications, being perhaps the most important the following ones:
particle acceleration takes place, which means that there are large amounts of energy contained in moving matter and magnetic fields in a low
entropic state, ready to be released in the form of heat and radiation; the acceleration efficiencies are high, pointing to quite extreme plasma
conditions from the point of view of particle confinement, flow velocities, and specific characteristics of radiation, matter, and magnetic fields
in the emitting region; the non-thermal energy outcome can be very large, which implies that the main radiative mechanism is efficient (e.g. close
to the saturation regime, when practically all the particle energy is radiated away). In addition, extended emission in the radio (e.g. Mirabel et
al. 1992) and X-ray bands (e.g. Corbel et al. 2002) has been detected, in some cases up to very large scales, implying that there is still
efficient non-thermal particle acceleration at the sites where this radiation is generated. 

Within the class of microquasars, the subclasses of low- and high-mass systems are distinguished depending on the mass of the stellar companion,
which can be smaller or bigger than the compact object mass. In general, high-mass microquasars have OB stars as companions, which produce strong
stellar winds and are very bright in the optical/UV band. Therefore, unlike low-mass microquasars, whose radiative processes would be determined
basically by the accretion/jet system only, high-mass microquasars should present a more complex phenomenology due to the presence of the wind and
the radiation field from the star. In such an environment the non-thermal radiation would be affected in different ways, being perhaps the most
relevant ones: dynamical interaction between the jet and the stellar wind (e.g. Perucho \& Bosch-Ramon 2008); attenuation of the radio emission due to
free-free absorption in the wind (e.g. Szostek \& Zdziarski 2007); inverse Compton (IC) scattering of relativistic electrons with stellar photons
(e.g. Paredes et al. 2000; Kaufman Bernad\'o et al. 2002); the stellar  wind could provide with targets for proton-proton collisions (e.g. Romero et al.
2003; Aharonian et al. 2006b); the gamma-rays of energy above the pair creation threshold would be absorbed via photon-photon interactions with the radiation field of
the star (Boettcher \& Dermer 2005); and the created secondary pairs would radiate under the magnetic and radiation fields present in the star
surroundings (Bosch-Ramon et al. 2008).

In this lecture, we try to give a broad overview of the physical processes involved in the formation and characterization of the observed non-thermal
emission from microquasars. In Section~\ref{scenario}, a qualitative and semi-quantitative description of the physical scenario is given, and the high
energy processes relevant in microquasars are introduced; in Sect.~\ref{radout}, the variability and spectral properties of  the emission are discussed;
finally, in Sect.~\ref{sum}, we conclude summarizing plus some comments and remarks.

\section{High energy processes in microquasars}\label{scenario}

\subsection{The physical scenario}\label{sce}

\begin{figure*}[p]  %%%%%%%%%%%%%%%%%%%%%%FIGURE 1ab %%%%%%%%%%%%%%%%%%%%%%%
\psfig{figure=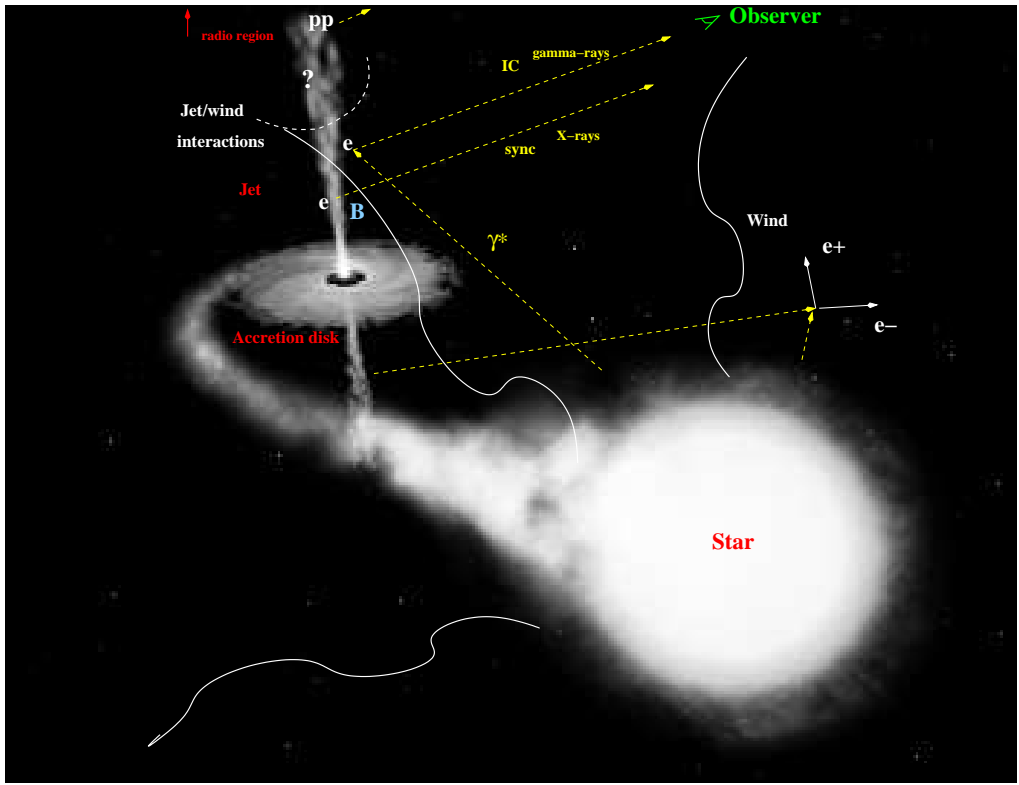,width=13.5cm}
%%          \hspace*{0.3cm}
%   \psfig{figure=fig1b.eps,width=6.3cm}       
\caption{Illustrative picture of the microquasar scenario, in which the main elements and processes 
considered in this work are shown (background image adapted from ESA, NASA, and F\'elix Mirabel -CEA, IAFE/CONICET-).\label{scen}}
\end{figure*}

In Fig.~\ref{scen}, we present the picture of a microquasar including the elements of the scenario: the star, the accretion disk\footnote{The
accreted matter has angular momentum, which leads to the formation of a disk. Some mechanism (e.g. turbulent viscosity -Shakura \& Sunyaev
1973- or an outflow -Bogovalov \& Kelner 2005-) can remove energy and angular momentum, allowing the material to drift towards regions of the
accretion disk closer to the compact object.}, the compact object, the jet, the magnetic field, a relativistic population of electrons and
protons in the jet, the stellar wind, the main radiative processes, and pair creation. Rather than going to details about the processes of generation of
jets, magnetic fields, acceleration of particles, and the like, we will give for granted what we already know from observations, i.e. that
microquasars present all these ingredients. Instead of that, what can be done is to constrain the models that 
describe these physical entities or processes using observations and basic tools from elementary physics. These tools
are presented in the next sections. 

\subsection{Energetics}

Non-thermal radio emission is produced in the jets of microquasars, but the energies of the photons produced in these jets likely go up to
gamma-rays (e.g. Paredes et al. 2000). Therefore, the first thing to do, in order to understand the microquasar non-thermal emission, is to look
for the origin of the energy powering these jets. Natural energy sources are the potential energy of the accreted matter (see sect.~3.1.2 in
Bosch-Ramon et al. 2006 for a semi-quantitative discussion), or the rotational energy of the compact object (e.g. Semenov et al. 2004; see also
Li et al. 2008). Since jet activity seems to be well correlated with accretion in microquasars (Fender et al. 2004), we will focus here on the
accretion energy requirements. The observed non-thermal luminosities can reach large values, $\sim 10^{35}-10^{36}$~erg~s$^{-1}$, and the jet
kinetic luminosity is, order of magnitude, similar or above these values, i.e. $\ga 10^{36}$~erg~s$^{-1}$. For an efficiency of the
accretion-ejection energy transfer of a 10\%, the required accretion energy budget will be $L_{\rm acc}\sim 10^{37}$~erg~s$^{-1}$. This value is
quite modest, being a few \% of the Eddington luminosity for a few M$_{\odot}$ compact object. In the phenomenological classification of Fender
et~al. (2004), this $L_{\rm acc}$-value would correspond to a microquasar in the low-hard state, when a persistent jet is expected to be present, i.e. 
microquasars can persistently power gamma-ray emission.

\subsection{Particle acceleration}\label{acc}

Presently, it is very difficult to distinguish between several mechanisms of particle acceleration in microquasar jets (for a discussion of
this, see Bosch-Ramon 2007 and references therein). Due to this fact, for the purpose of this lecture, 
it is more convenient to investigate the acceleration of particles
using a simple approach. The first step is to see whether a necessary condition 
for the acceleration of a charged particle with certain energy is fulfilled,
which is given by the Hillas criterium (Hillas 1984). This consists on the fact that particles can only be accelerated if their
Larmor radius ($r_{\rm L}=E/qB_{\rm a}$; where $B_{\rm a}$ is the accelerator magnetic field, and $q$ and $E$ are the charge and energy of
the particle, respectively) is smaller than the accelerator size ($l_{\rm a}$); otherwise particles escape the accelerator. This
limits the highest achievable energy to (if not specifically stated, the units in this work are cgs):
\begin{equation}
E<qB_{\rm a}l_{\rm a}\,.
\label{larm}
\end{equation}
Making a step further, to determine whether particles can be accelerated up to a certain energy, the specific acceleration and energy loss 
(or particle escape) mechanisms are to be known: $t_{\rm acc}=t_{\rm cool/esc}$\,. 
In general, the acceleration timescale can be expressed as:
\begin{equation}
t_{\rm acc}=\eta{r_{\rm L} \over c}=\eta \frac{E}{qBc}\,,
\label{accrate}
\end{equation}
where $\eta$ is a dimensionless phenomenological parameter (or function) representing the acceleration efficiency, different in each 
acceleration scenario. The particular case of $\eta=1$ corresponds to the shortest possible acceleration time
independently of the acceleration mechanism. An instance for $\eta$ can be given for the case of
non-relativistic diffusive shock acceleration (plane shock with weak magnetic field, in the test particle approximation -Drury 1983-):
\begin{equation}
\eta=2\pi{D\over D_{\rm Bohm}}\left(c\over V_{\rm sh}\right)^2\,,
\label{accef}
\end{equation}
where $ V_{\rm sh}$ is the shock velocity, and $D$ is the  diffusion coefficient ($D_{\rm Bohm}$ in the Bohm limit). For $V_{\rm
sh}=3\times 10^9$~cm~s$^{-1}$ and $D=D_{\rm Bohm}$, $\eta\sim 10^3$. 

Without focusing on any particular acceleration  mechanism, it is worthy noting that the final product of particle acceleration  of a diffusive or stochastic mechanism will
be a power-law ($Q(E)\propto E^{-h}$), and $h\sim 2$ is commonly adopted. Diffusive non-relativistic strong shock acceleration results in $h=2$. Another point to notice is
that, usually,  the particles in the emitter, e.g. the jet, are considered to have an isotropic velocity distribution 
in the jet reference frame (RF), due to  deflection in the
randomic component of the magnetic field, which moves solidary with the jet matter.

\subsection{Particle propagation}

In a medium of diffusion coefficient $D$, particles propagate in one particular direction with the time dependence $l_{\rm diff}\approx\sqrt{2Dt}$.
Under the impact of cooling, 
the typical distance that particles of energy $E_{\rm TeV}=E/1~{\rm TeV}$, and under a magnetic field of $B_{\rm
G}=B/1~{\rm G}$, can reach is:
\begin{equation} 
l_{\rm diff}\approx 10^{10}\,E_{\rm TeV}^{1/2} B_{\rm G}^{-1/2}t_{\rm cool}^{1/2}
\left(\frac{D}{D_{\rm Bohm}}\right)^{1/2}\,\,{\rm cm}\,,
\label{diffs}
\end{equation}
where the Bohm rate is the slowest possible diffusion rate, $D_{\rm Bohm}=r_{\rm L}c/3$, and $t_{\rm cool}$ 
is the cooling timescale in seconds of the dominant loss mechanism. If the medium in which particles are embedded is also moving (e.g. a jet), there 
is in addition advective transport:
\begin{equation} 
l_{\rm adv}\approx 10^{10}\left(\frac{V_{\rm adv}}{10^{10}\,{\rm cm~s}^{-1}}\right)t_{\rm cool}\,\,{\rm cm}\,.
\label{advs}
\end{equation}
These formulae allow us to estimate, under different cooling processes, whether particles can propagate significantly 
when comparing these distances with the accelerator size. For instance, jet advection may transport particles up to
large distances. If the radiative processes were still efficient, such a situation would imply that the emitter 
size $l\gg l_{\rm a}$ (for a thorough discussion, 
see Khangulyan et al. 2008).

\subsection{Cooling processes}\label{loss}

We consider here the following non-thermal processes: relativistic Bremsstrahlung, 
synchrotron and IC (for an exhaustive review, see Blumenthal \& Gould 1970), when leptons are the main emitters; and proton-proton interactions (see Kelner et al.
2006), when protons are the particles that produce the emission to study. Here we briefly list the relevant cooling timescales of these
processes. The adiabatic loss time is also given, since relativistic particles could lose energy exerting work on the medium.

Relativistic electrons radiate Bremsstrahlung under the effect of the electric field in the surroundings of an atom. 
The characteristic cooling timescale is (e.g. Cheng \& Romero 2004):
\begin{equation}
t_{\rm Br}\sim \frac{10^{15}~{\rm cm}^{-3}}{n}\,{\rm s}\,,
\label{tsync}
\end{equation}
where $n$ is the density of the medium. Synchrotron radiation originates when an electron moves in an irregular magnetic field, spiraling
around the chaotically oriented magnetic lines. The electron suffers Lorentz forces and radiate in the direction of motion the energy corresponding to the 
momentum perpendicular to the magnetic field. The typical
frequency of the outgoing synchrotron photon is $\nu=6.27\times 10^{18}BE^2$. It is thought that a significant part of the magnetic energy
density in jets is in the form of a chaotic magnetic field (e.g. Biermann \& Strittmater 1987) attached to the
plasma. This random magnetic field also confines the relativistic particles, which diffuse inside the jet instead of escaping from it almost at the speed of
light. The characteristic energy loss timescale for synchrotron emission is:
\begin{equation}
t_{\rm sy}\approx 4\times 10^2 B_{\rm G}^{-2} E_{\rm TeV}^{-1}\,{\rm s}\,.
\label{tsync}
\end{equation}
Under the presence of radiation fields, relativistic electrons also interact with the ambient photons via IC scattering, 
in which the scattered photon has increased 
its energy by $4(E/m_{\rm e}c^2)^2$ and takes the direction of the scattering electron.
If the energy of the target photon ($\epsilon_0$), in the electron
RF, is $< m_{\rm e}c^2$, the interaction takes place in the classical Thomson regime; in the emitter RF, the electron energy must be 
$< m_{\rm e}c^2/\epsilon_0$. The associated characteristic timescale is:
\begin{equation}
t_{\rm IC~T}\approx \frac{16}{u_{\rm rad}E_{\rm TeV}}\,{\rm s}\,,
\label{ticth}
\end{equation}
where $u_{\rm rad}$ is the target photon field energy density (cgs).
In case the target photon, in the electron RF, has an energy $> m_{\rm e}c^2$, the interaction enters in the Klein-Nishina regime, 
with a characteristic timescale (Khangulyan et al. 2008):
\begin{equation}
t_{\rm IC~KN}\approx 10^4 (u_{\rm rad})^{-1} \left(\epsilon_0\over 10\,{\rm eV}\right)^{1.7}\, E_{\rm TeV}^{0.7}\,{\rm s}\,.
\label{tickn}
\end{equation}
It is worthy noting here that IC scattering has a strong dependence on the angle of interaction between the incoming electron which becomes 
less important when entering in the Klein-Nishina regime. This has to be accounted when computing the radiation for a specific system and orbital phase, since the geometry of
the interaction changes along the orbit (see Khangulyan et al. 2008 for a deeper discussion on  this; see Blumenthal \& Gould 1970 for the
angle-averaged cross section, and Bogovalov \& Aharonian 2000\footnote{Eq.~(19) in this work lacks a $(1-\cos(\theta))$ inside the integral
($\theta$ is the angle between the electron and photon directions of motion in the laboratory RF).} for the angular dependent one).  We note however that the particle
distribution is usually assumed isotropic in the emitter RF and $t_{\rm IC}$ will not depend on the angle. This emitter may be moving in a 
certain direction with respect to the observer, e.g. in the jet, in which case Doppler boosting has to be taken into account.

If relativistic protons are present, they will lose energy interacting with the ambient atoms (e.g. Romero et al. 2003;
Aharonian et al. 2006b). At the relevant energies, i.e. $>$~GeV,
ionization losses become negligible, and proton-proton collisions are the dominant cooling channel with timescales:
\begin{equation}
 t_{\rm pp}\approx {10^{15}\,{\rm cm}^{-3}\over n_{\rm t}}\, {\rm s}\,,
\label{tpp}
\end{equation}
where $n_{\rm t}$ is the density of targets (cgs).
After colliding, among other possible interaction products, about half of the proton energy goes to $\pi^{0/\pm}$, which decay to
gamma-rays/$\mu^{\pm}$/$\nu_{\mu}$,  and $\mu^{\pm}$ decay to $e^{\pm}$ (see Orellana et al. 2007 for their radiative relevance 
in microquasars), $\nu_{\rm e}$ and $\nu_{\mu}$. Roughly, 1/6 of the proton energy
goes to gamma-rays, a similar amount to $\nu$,  and 1/12 to $e^{\pm}$. If dense and very hot photon fields and ultrarelativistic
nuclei were present, photo-meson production and photo-disintegration could play some role, and even synchrotron proton radiation may be
efficient under certain conditions (Vila \& Romero 2008). Here we will not consider them for simplicity, but also because they are in general
little efficient or start to work
at energies that may not be reached due to acceleration constraints (Bosch-Ramon \& Khangulyan 2008, in preparation).

The cooling timescale for particles exerting work on the surrounding medium is:
\begin{equation}
t_{\rm ad}\approx \frac{2l}{3V_{\rm exp}}\, {\rm s}\,.
\label{tad}
\end{equation}
This can be considered to happen in case the relativistic particles are confined within a medium that suffers expansion, 
like a jet in overpressure with its
environment.

\subsection{Photon attenuation processes}\label{att}

Absorption of gamma-rays with energies above the pair creation threshold ($\epsilon_{\rm th}=2m_{\rm e}^2c^4/\epsilon_0(1-\cos(\theta))$; where 
$\epsilon_0$ is the target photon energy and $\theta$ is the interaction angle between the directions
of the VHE and the stellar photons) will occur if
there is a photon field surrounding the VHE emitter. In case of microquasars harboring massive stars, the opacity for one photon to be
absorbed, related to the interaction probability, 
can be close to or much larger than 1, i.e. photon-photon absorption is either optically thin or optically 
thick. To estimate the significance of this process, we can use the next simple expression:
\begin{equation}
\tau_{\gamma\gamma}\approx 10^{-25}\frac{u_{\rm rad}R_{\rm orb}}{\epsilon_*}\,,
\end{equation}
where typically the system size $R_{\rm orb}\sim 10^{12}-10^{13}$~cm, $\epsilon_*\sim 1.6\times 10^{-11}$~erg ($T_*\approx 4\times 10^4$~K) is the typical energy
of the stellar photons, and $u_{\rm rad}\sim 100$~erg~cm$^{-3}$. $\tau_{\gamma\gamma}$ is the opacity for gamma-rays 
with energies $\approx
3.7\times \epsilon_{\rm th}$ in an isotropic target photon field (when the cross section is the largest; see Coppi \& Blandford 1990, eq.~4.7).

Actually, in most high-mass microquasars (LS~5039, Khangulyan et al. 2008;
Cygnus~X-1, Bednarek \& Giovanelli 2007; Cygnus~X-3, Protheroe \& 
Stanev 1987;  SS~433, Reynoso et al. 2008; and LS~I~+61~303, Romero et al. 2007)
opacities significantly above 1 will occur. The occurrence of absorption is important not only because it reduces the amount of VHE photons
that escape the system, but also because it depends on $\theta$. 
This dependence is present through the interaction probability and $\epsilon_{\rm th}$. 
In microquasars, since the interaction angle changes along the orbit, as well as the surface density of target
photons seen by the observer, opacity changes along the orbit (a discussion of this can be found in Khangulyan et al. 2008; the angle
averaged pair creation rate can be found in Coppi \& Blandford 1990, and the angle dependent cross section in Gould \& Schr\'eder
1967)\footnote{When computing the opacity, the factor $(1-\cos(\theta))$ should be 
also considered in the calculation, as in the case of 
angle-dependent IC scattering.}.

Electromagnetic cascade can take place efficiently when KN IC losses are dominant and
$\tau_{\gamma\gamma}\gg 1$. The former implies that the ambient magnetic field outside the emitter must be
low enough, i.e.:
\begin{equation}
B_{\rm c}<10\left(L_*\over 10^{39} {\rm  erg~s}^{-1}\right)^{1/2}\left(R\over R_{\star}\right)^{-1}{\rm G}\,. 
\label{eq:b_crit}
\end{equation}
At $B_{\rm c}$, KN IC losses are equal to synchrotron losses for 1~TeV photons. Electromagnetic cascading consist 
on a VHE electron that creates a VHE photon via IC, which is
absorbed in the stellar field creating a VHE electron/positron pair\footnote{Well above the threshold, when cascading is efficient -deep KN IC-, one member of the pair
will take most of the initial photon energy (e.g. Boettcher \& Schlickeiser 1997).}, 
which to its turn creates another VHE photon via IC, etcetera (this is quite constraining, actually, since $B_{\rm a}$
could be significantly larger than $B_{\rm c}$, as noted in Sect.~\ref{targets}). 
Electromagnetic cascading occurring deep in the KN regime reduces effectively
the opacity of the system. If the magnetic field were similar to or larger than $B_{\rm c}$, the absorbed energy would be mainly released via synchrotron
radiation (see Bosch-Ramon et al. 2008).

Beside gamma-ray absorption due to pair creation, it is worthy also to note that radio emission can be attenuated due to free-free
absorption in the stellar wind. An estimate for the free-free opacity (Rybicki \& Lightman 1979) can be obtained from the next formula:
\begin{equation}
\tau_{\nu~ff}\la 40\frac{\dot{M}_7^2}{\nu_{\rm 5~GHz}^2R_{\rm 0.3~AU}^3V_{\rm w~8.3}^2T_{\rm w~4}^{3/2}}\,,
\end{equation}
where $\dot{M}_7=\dot{M}/10^{-7}~$M$_{\odot}$~yr$^{-1}$ is the stellar mass loss rate, $\nu_{\rm 5~GHz}=\nu/5~{\rm GHz}$ the frequency,
$R_{0.3}=R/0.3~{\rm AU}$ the distance to the star, $V_{\rm w~8.3}=V_{\rm w}/2\times 10^8$~cm~s$^{-1}$ the stellar wind velocity, and $T_{\rm w~4}=T_{\rm w}/10^4$~K the wind
temperature. This estimate could be affected by the real structure of the stellar wind, which may be clumpy (e.g. Owocki \& Cohen 2006), reduced or increased 
depending on the clump mass density and number.
This estimate suggests that the observed radio emission is produced outside the binary system in high-mass microquasars ($\ga
AU$), at spatial scales that may be only marginally resolvable by the present radio interferometer instruments, but still far from the jet ejection
region. 

Another source of radiation attenuation is synchrotron self-absorption. 
Electrons of certain energies, within a source of certain size and magnetic
field, can efficiently absorb the synchrotron emission produced by them. For an homogeneous emitter with a population of relativistic
electrons producing synchrotron emission, the synchrotron self-absorption frequency is (see chapter 3 in Pacholczyk 1970): 
\begin{equation}
\nu_{\rm ssa}\approx 2\,c_1\,(lc_2K)^{2/(p+4)}\,B^{(p+2)/(p+4)}\,{\rm GHz}\,,
\end{equation}
where $K=N(E)/E^{-p}$ ($p$ here is the one at the relevant electron energies) is the electron energy distribution normalization\footnote{It is worthy noting that the injection 
electron spectrum $Q(E)$ and the particle energy distribution $N(E)$ are different functions. The first one gives the energy dependence of the injected 
particles, and the second one gives the same but for the final particles that suffered the different cooling and escape processes that take place
in the emitter. The units may not coincide.}, 
$c_1=6.3\times 10^{18}$, $c_2\approx 10^{-40}$ ($c_2$ actually depends on $p$, but the given value works for 
$p\approx 2-2.5$), and $l$ is the size of the
emitter. We notice that, if the emitter has a complex structure or is strongly inhomogeneous, $\nu_{\rm ssa}$ may be significantly different.

Finally, ionization of the stellar wind atoms can absorb the lower energy part of the X-ray spectrum, and also induce relativistic particles cooling.
We do not discuss these processes here though note that they may be relevant in some specific situations (e.g. Bosch-Ramon et al. 2007; 2008).

\subsection{Targets for radiation and energy losses}\label{targets}

Relativistic particles interact with different targets: 
magnetic, photon and matter fields, producing emission and subsequently losing energy. In microquasars,
there are different sources for the mentioned targets. Here, we will focus on those related to the star (for the case of high-mass systems),
the jet, and the environment at large scales. In the regions close to the compact object, where the jet is thought to be produced, dense
target fields from the accretion disk could be present as well, but given our lack of knowledge concerning the real jet and environment
properties in those regions, we do not consider them here (for a brief discussion, see Bosch-Ramon 2007). 

The star is a very important ingredient in high-mass microquasars. 
On one hand, the stellar photon field, with typical luminosities $L_*\sim 10^{38}-10^{39}$~erg~s$^{-1}$ and
temperatures $\sim 4\times 10^4$~K, renders stellar IC a very efficient process, as well as photon-photon absorption. On the other hand, the
stellar wind provides targets for proton-proton collisions, could be a source of radio and soft X-ray radiation attenuation, and may play a 
dominant role on the jet dynamics at spatial scales $\sim R_{\rm orb}$ (not treated here; see Perucho \& Bosch-Ramon 2008 for a thorough discussion). 
The wind can have a very complex structure as noted in Sect.~\ref{att}, although we estimate its density 
assuming that this is an homogeneous supersonic radial flow of velocity $\sim
2\times 10^8$~cm~s$^{-1}$:
\begin{equation}
n_{\rm w}=2\times 10^8\,
\left(\frac{\dot{M}_*}{10^{-7}~{\rm M}_{\odot}~{\rm s}^{-1}}\right)
\left(\frac{3\times 10^{12}~{\rm cm}}{R_{\rm orb}}\right)^2\,{\rm cm}^{-3}\,. 
\end{equation}
Concerning the stellar photon field, the photon number density (valid for a black body)
can be roughly approximated by:
\begin{equation}
n_{\rm ph}=2\times 10^{13}\,
\left(\frac{L_*}{10^{39}~{\rm erg}~{\rm s}^{-1}}\right)
\left(\frac{3\times 10^{12}~{\rm cm}}{R_{\rm orb}}\right)^2\,{\rm cm}^{-3}\,.
\end{equation}
For the magnetic field in the surroundings of the star, we can adopt a $R$-dependence 
of $B_{\rm a}$ in the range $r=1-3$ (Usov et al. 1992), which would correspond to a toroidal--radial--dipolar dependence normalized to 
$B_*$ at the stellar surface ($R_*$):
\begin{equation}
B_{\rm a}=100\,
\left(\frac{B_*}{100~{\rm G}}\right)
\left(\frac{R_*}{R}\right)^{r}\,{\rm G}\,.
\end{equation}
It is worthy noting that $B_*$ is typically 100--1000~G in OB stars, found for instance using the Zeeman effect
(Donati et al. 2002). Further evidences of the presence of the magnetic field come from the detection of non-thermal radio emission 
from isolated massive stars (e.g. Benaglia 2005 and references there in), and X-ray observations (e.g. Stelzer et al. 2005). Recalling what was said in 
Sect.~\ref{att} regarding electromagnetic cascades, the typical $B_*$-values would seem to preclude the occurrence of efficient cascading in high-mass microquasars.

The magnetic field in the jet could be assumed to go like $1/z$ (e.g. for a toroidal magnetic field), where $z$ is the jet height, and normalized to the magnetic field value at
some height (i.e. $B_0$ at $z_0$):
\begin{equation}
B_{\rm jet}=1\,\left(\frac{B_0}{10^5~{\rm G}}\right)\left(\frac{10^{8}~{\rm cm}}{Z_0}\right)\left(\frac{10^{12}~{\rm cm}}{Z}\right)\,{\rm G}\,.
\end{equation}
The presence of magnetic field and relativistic electrons in the jet leads to synchrotron emission. This process generates a target photon field inside the jet 
that could be 
suitable, under certain conditions of jet radius ($R_{\rm jet}$), $B_{\rm jet}$, and injected particle luminosity ($L_{\rm rel}$), 
for efficient synchrotron self-Compton (SSC), i.e. 
scattering of the synchrotron photons by the relativistic electrons that produced them.
A relationship that tells when synchrotron self-Compton losses start to dominate over synchrotron ones is the following:
\begin{equation}
u_{\rm B}=\frac{B_{\rm jet}^2}{8\pi}\sim\frac{L_{\rm sync}}{\pi R_{\rm jet}^2 c}\la\frac{L_{\rm rel}}{\pi R_{\rm jet}^2 c}\,,
\end{equation}
where the main assumption is that the synchrotron emission (of luminosity $L_{\rm sync}$) 
comes mostly from one compact region of magnetic field $B_{\rm jet}$. 
The treatment of SSC losses, when dominant, is very complicated since it is non-linear, i.e. the radiation
and the particle distribution are strongly coupled, with feedback effects. Therefore, it is important to know when SSC losses are to be accounted for (e.g.
Bosch-Ramon \& Paredes 2004).
Actually, in some cases the SSC radiation can be computed without the need 
of taking into account its losses since other energy losses, like synchrotron itself, or external IC, are dominant.
Finally, the matter density of a conical supersonic jet can be expressed in the next form:
\begin{equation}
n_{\rm jet}=2\times 10^{15}\,
\left(\frac{L_{\rm jet}}{(\Gamma_{\rm jet}-1)\,10^{37}~{\rm erg}~{\rm s}^{-1}}\right)
\left(\frac{10^{10}~{\rm cm}~{\rm s}^{-1}}{V_{\rm jet}}\right)
\left(\frac{10^7~{\rm cm}}{R_{\rm jet}}\right)^2\,{\rm cm}^{-3}\,,
\end{equation}
where $V_{\rm jet}$ is the jet speed and $\Gamma_{\rm jet}$ the jet Lorentz factor. 
In general, densities are expected to be low, probably not enough to produce gamma-ray emission via relativistic Bremsstrahlung, 
or proton-proton interactions (regarding the latter, see otherwise the case of SS~433, which has very heavy jets -Reynoso et al. 2008-).

At the largest scales, where the microquasar jet terminates, the treatment of the ambient radiation, matter and magnetic fields becomes more
complex. For instance, in massive systems, a powerful stellar wind would reduce the density and increase the pressure of the jet environment via
shocking the ISM, modifying the whole dynamics of the jet/medium interaction and introducing inhomogeneities in the surroundings. Also, high-mass
microquasars, being {\it young} objects, tend to be embedded in a dense medium, unlike low-mass microquasars, which can be quite {\it old} and
present even in the halo. In the case of low-mass systems, the environment of the microquasar may have very different densities,
$10^{-4}$--$\,>1$~cm$^{-3}$, depending on the location within the Galaxy, and the ISM, which could present by itself strong inhomogeneities at pc scales. In
addition, a massive and hot star could generate a radiation field above the local one (i.e. the galactic background). Otherwise, the latter would
be the only possible target for IC interactions, which would typically yield quite low efficiencies. Like the density, the magnetic field in the
jet termination regions is an uncertain parameter, but some equipartition arguments and simple jet/medium interaction models give values $\sim
10^{-5}-10^{-3}$~G, as found e.g. by Bordas et al. 2008. In that work, the authors show that the non-thermal emission from the jet termination
regions could be detectable\footnote{Interestingly, microquasars are not the only galactic sources that produce non-thermal emission via
jet-environment interactions (see Araudo et al. 2007 and references therein).}.

\section{On the radiative output}\label{radout}

In this section, we present some examples of the spectra and the lightcurves that the emission from microquasars could show. In order to {\it shape} the spectrum and 
the lightcurve,
several possible processes to explain observations at different wavelengths are considered. First, a semi-qualitative/quantitative insight into the formation of the particle
energy distribution is given, as well as a brief description of the shape of the spectrum of the radiation produced from different particle energy distributions. 

\subsection{The energy distribution of particles and the radiation spectrum}

\begin{figure*}[p]  %%%%%%%%%%%%%%%%%%%%%%FIGURE 1ab %%%%%%%%%%%%%%%%%%%%%%%
\psfig{figure=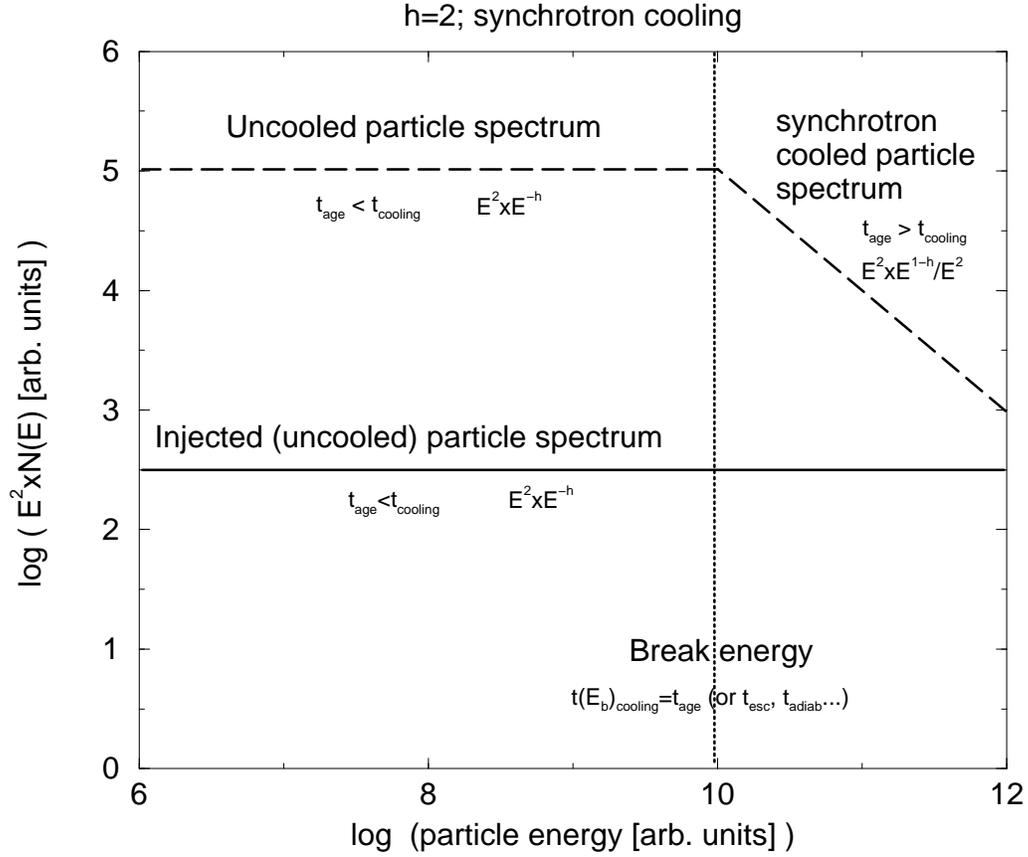,width=13.5cm}
%%          \hspace*{0.3cm}
%   \psfig{figure=fig1b.eps,width=6.3cm}       
\caption{The sketch of a particle energy distribution times $E^2$ is shown. The electron population suffers synchrotron losses and the cooled 
particle distribution is $\propto E^2\times E^{-3}$ for h=2. In the case shown here, below the break energy the age of the source equals or is smaller than the 
synchrotron energy loss timescale, and the spectrum is still uncooled. \label{part}}
\end{figure*}

In Section~\ref{acc}, we have already mentioned that the injected particle spectrum is usually adopted with index $h=2$. Although the complexity of
the emitter (radiative losses, non radiative losses, escape losses, etc.) may require the use of a very complex model setup, the basic
features of the evolved particle energy distribution can be extracted from a very simple case: a (one-zone) homogeneous emitter with particle
escape in the steady regime (see Eq.~3--7 in Khangulyan et al. 2007). The main idea is that the final particle energy distribution,
$N(E)$\footnote{The emitter is considered steady and homogeneous, without spatial nor time dependences in $N(E)$.}, is: 
\begin{equation}
\propto E^{-p}=
E^{-h+1}/\dot{E}{\rm ~~>\,minimum~injection~energy}\,, \propto 1/\dot{E}{\rm ~~otherwise}\,.
\end{equation}
From $\dot{E}=-E/t_{\rm cool}\propto E^g$, we can roughly give the $N(E)$ shape for different processes, where $g=2$ in the case of synchrotron
and Thomson IC, $g\sim 0$ in the case of KN IC and ionization losses, $g\sim 1$ in the case of relativistic Bremsstrahlung, $g\sim 1$ in the case
of proton-proton interactions, and $g=1$ in the case of adiabatic losses. If an escape time is introduced (e.g. due to convection by the outflow
away from the shock that accelerates particles), no particles {\it older} than this escape time will be present, and they will not contribute to
the particle energy distribution below the energy at which they escape. The particles that do not have time to cool will escape the source at the
same rate whatever their energy. It implies that, in general, below the escape energy, $N(E)\propto E^{-h}$ (above the minimum injection
energy) and $=0$ (below the minimum injection energy). Analogously, in the time dependent case,  i.e the age of the accelerator < the cooling
timescale at a certain energy, the particle energy distribution has $p=h$ below this energy (so-called break energy). All this is valid for
protons and for electrons. In Fig.~\ref{part}, the energy distribution of electrons evolving under synchrotron cooling is shown to illustrate the
effect of energy losses in the injected particle spectrum.

For the shape of the spectral energy distribution ($\nu F(\nu)$, where $F(\nu)$ is the specific flux), it can be 
written:
\begin{equation}
\nu F( \nu )=\nu^2\,n(\nu)=\nu\int_{E_{\rm min(\nu)}}^{E_{\rm max}} N(E)P(E,\nu)dE\approx \nu\dot{E}_{\rm rad}(E(\nu))N(E(\nu))dE(\nu)/d\nu\,,
\end{equation}
in the delta function approximation for the specific power $P(E,\nu)$ of one particle of energy $E$, 
approximately
valid for the radiative processes presented in Sect.~\ref{loss}. $n(\nu)$ is the specific photon number. 
The cooling and the radiation rates, $\dot{E}_{\rm cool}$ and 
$\dot{E}_{\rm rad}$, refer to the radiation we calculate and the dominant cooling mechanism affecting particles, respectively. $\dot{E}_{\rm rad}$ and 
$\dot{E}_{\rm cool}$ do not necessarily have to be the same.

\begin{figure*}[p]  %%%%%%%%%%%%%%%%%%%%%%FIGURE 1ab %%%%%%%%%%%%%%%%%%%%%%%
\psfig{figure=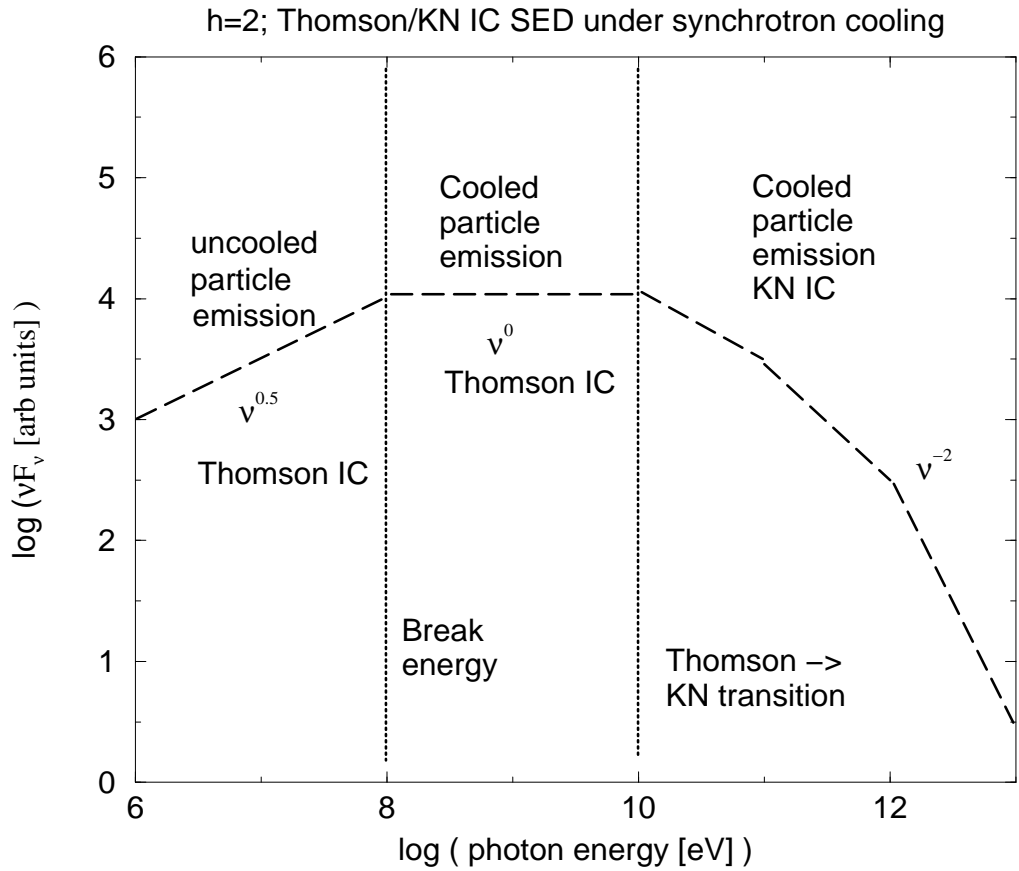,width=13.5cm}
%%          \hspace*{0.3cm}
%   \psfig{figure=fig1b.eps,width=6.3cm}       
\caption{The sketch of a spectral energy distribution of IC emission produced by electrons cooled by synchrotron emission. The synchrotron component is not 
shown. \label{sed}}
\end{figure*}

To go further, $E(\nu)$ is required, and a dependence of the kind $E(\nu)\propto \nu^l$ will be adopted, which is roughly correct for the mechanisms considered here.
For synchrotron and Thomson
IC, $l=1/2$; for KN IC, relativistic Bremsstrahlung and proton proton interactions, $l\approx 1$. Therefore, for a cooled particle population,
\begin{equation}
\nu F( \nu )\propto \nu\dot{E}_{\rm rad}(E(\nu))E(\nu)^{1-h}\nu^{l-1}/\dot{E}_{\rm cool}(E(\nu))\propto \nu^{l(g_{\rm rad}-g_{\rm cool}+2-h)} \,,
\end{equation}
which implies, for instance, that $\nu F(\nu)\propto$ constant for synchrotron, Thomson IC and proton proton radiation under $\dot{E}_{\rm rad}=\dot{E}_{\rm cool}$, 
or $\nu^{-2}$ for KN IC under synchrotron cooling. This second case is presented in Fig.~\ref{sed}. For an uncooled particle population,
\begin{equation}
\nu F( \nu )\propto \nu\dot{E}_{\rm rad}(E(\nu))E(\nu)^{-h}\nu^{l-1}\propto \nu^{l(g_{\rm rad}+1-h)}\,,
\end{equation}
and below the minimum injection energy (if particles had time to cool before escaping or at the age of the source):
\begin{equation}
\nu F( \nu )\propto \nu\dot{E}_{\rm rad}(E(\nu))\nu^{l-1}/\dot{E}_{\rm cool}\propto \nu^{l(g_{\rm rad}-g_{\rm cool}+1)}\,.
\end{equation}

Below the escape energy and the minimum injection energy, the delta function approximation does not work, and the proper emissivity function is to be accounted for (e.g. for
synchrotron/Thomson IC emission: $\nu F(\nu)\propto \nu^{4/3}$; for relativistic Bremsstrahlung: $\nu F(\nu)\propto \nu^1$). Complex shapes for the particle spectrum, very
different from a power-law, and the presence of low energy and high energy cut-offs or extreme slopes would modify the simple schema presented here.

\subsection{Variability and dominant emission mechanisms at different wavelengths}

In this section, we describe briefly the variability of the source depending on the energy range. For simplicity, we keep our discussion in the context of one-zone or homogeneous
models. We note that propagation effects, which can lead to emitter inhomogeneity, can change the variability pattern of the radiation, e.g. smoothing it (see Fig.~22 in
Khangulyan et al. 2008 for an example of the impact of particle advection in the lightcurve and spectrum), or increasing it via particle escape (e.g. Khangulyan et al. 2007). 

In the simple context adopted here, e.g. one-zone model, different factors can influence on the emission behavior, like the particle injection rate, the magnetic field, the target
photon density, the non-radiative cooling rate, the escape timescale, and the geometry changes of the binary/observer system along the orbit. It is worthy noting that in the
saturation regime, the change of target densities does not {\bf necessarily} render variability as long as the injected luminosity is constant (e.g. changing the stellar photon
density along the orbit in an eccentric system does not imply modulation of the IC emission).

{\bf Radio}

Radio emission, of synchrotron origin in microquasars, can be affected by all the mentioned factors that influence on the lightcurve but the system geometry. Nevertheless, it
is worth to note that synchrotron self-absorption makes the emitter to become opaque to its own synchrotron radiation below a certain frequency. 
In such a regime of emission, variations in the observed
spectral break frequency ($\nu_{\rm ssa}$; the optically thick-optically thin transition) can occur depending from which direction the observer sees an (e.g. asimetric)
emitter. We also note that radio photons are subject to free free absorption, very likely to occur in the stellar wind. This will also introduce orbital dependent
variability.

{\bf X-rays}

X-rays could be of synchrotron origin, produced by the high energy part of $N(E)$, or IC origin, produced by the low energy part of $N(E)$ (i.e. in the Thomson regime). In the
former case, the radiation would be affected by all the factors mentioned above but changes of geometry, unless the electron distribution and/or the magnetic field configuration
are themselves anisotropic. Synchrotron self-absorption is hardly relevant at these energies. IC can be affected, together with the other  variability sources, by geometrical
effects via the next dependence: $\nu F(\nu)\propto (1-\cos(\theta))$, since $(\theta)$ will change strongly along the orbit since the stellar photons come from the star
direction, which is orbital phase dependent from the observer point of view. Relativistic Bremsstrahlung can be hardly efficient in this energy range, since it is very hard below
the energies around the electron rest mass one ($\sim$~MeV) and will fall probably below the synchrotron and/or IC fluxes. X-ray attenuation due to ionization of wind atoms, with
a changing hydrogen column density along the orbit, may be a source of variability.

{\bf Gamma-rays (GeV)}

The emission in this energy range could be due to relativistic Bremsstrahlung, proton-proton interactions, or IC scattering. If hadronic and the leptonic acceleration were equally
efficient, and IC scattering negligible (e.g. under a low-mass companion photon field), relativistic Bremsstrahlung would dominate due to the smaller energy transferring in the
case of (proton-proton) proton to gamma-rays compared with (relativistic Bremsstrahlung) electron to gamma-rays. Nevertheless, very high densities would be required (see
Sect.~\ref{loss} and \ref{targets}). If a massive and hot primary is present, stellar (Thomson) IC will be a much more efficient process. SSC radiation could be still relevant for
the case of a high-mass microquasar, and certainly is a good candidate for low-mass microquasars\footnote{There could be additional IC components due to the presence of strong
accretion disk and/or corona photon fields (e.g. Bosch-Ramon et al. 2006).}. As noted above, powerful jets containing relativistic protons in low-mass microquasars may also
present significant hadronic radiation from other channels.

The variability in this energy range could be due to all the effects mentioned above. In particular, the angular dependence of (Thomson) IC ($\sim (1-\cos(\theta))$), implies that
less IC emission goes to the observer when the emitter is in front of the star, receiving the stellar photons from behind, thought to happen during the superior conjunction of the
compact object. In addition, the IC emission from secondary pairs or the occurrence of electromagnetic cascading, reprocessing energy above $\sim 100$~GeV down to lower energies,
may affect the GeV variability as well, since the observable outcome of cascading is sensitive to the binary/observer geometry. 

{\bf Gamma-rays (TeV)}

The radiation processes contributing at these energies are the same as those contributing in the GeV range. Nevertheless, depending on the particle energy distribution and cross
section behavior at high energies, some processes may be more or less significant in the GeV and TeV regimes. The nature of the emitting particles are also relevant. For instance,
a synchrotron dominated particle energy distribution is much softer than a KN IC one, and electrons may be softer or harder than protons depending on the cooling conditions,
although protons will in general yield a harder radiation spectrum at the highest energies for the same $h$.

The variability at these energies can be produced by the same factors as in the GeV range. Regarding the angle-dependence of IC scattering, at TeV energies this process takes
place in the KN regime. There is still some modulation of the emission, similar to some extent to $(1-\cos(\theta))$, but the deeper in the KN regime the IC scattering occurs, the
smaller the modulation. At energies $\gg$~TeV, there is basically no modulation. The changing angular dependence at different energies renders a complex spectral time behavior.
There is additional modulation due to photon-photon absorption above the pair creation threshold, due to $\sim (1-\cos(\theta))$ and electromagnetic cascading, and the pair
creation threshold also depends on the photon-photon scattering angle (recall: $\epsilon_{\rm th}=2m_{\rm e}^2c^4/(1-\cos(\theta))\epsilon_0$). 

\begin{figure*}[p]  %%%%%%%%%%%%%%%%%%%%%%FIGURE 1ab %%%%%%%%%%%%%%%%%%%%%%%
\psfig{figure=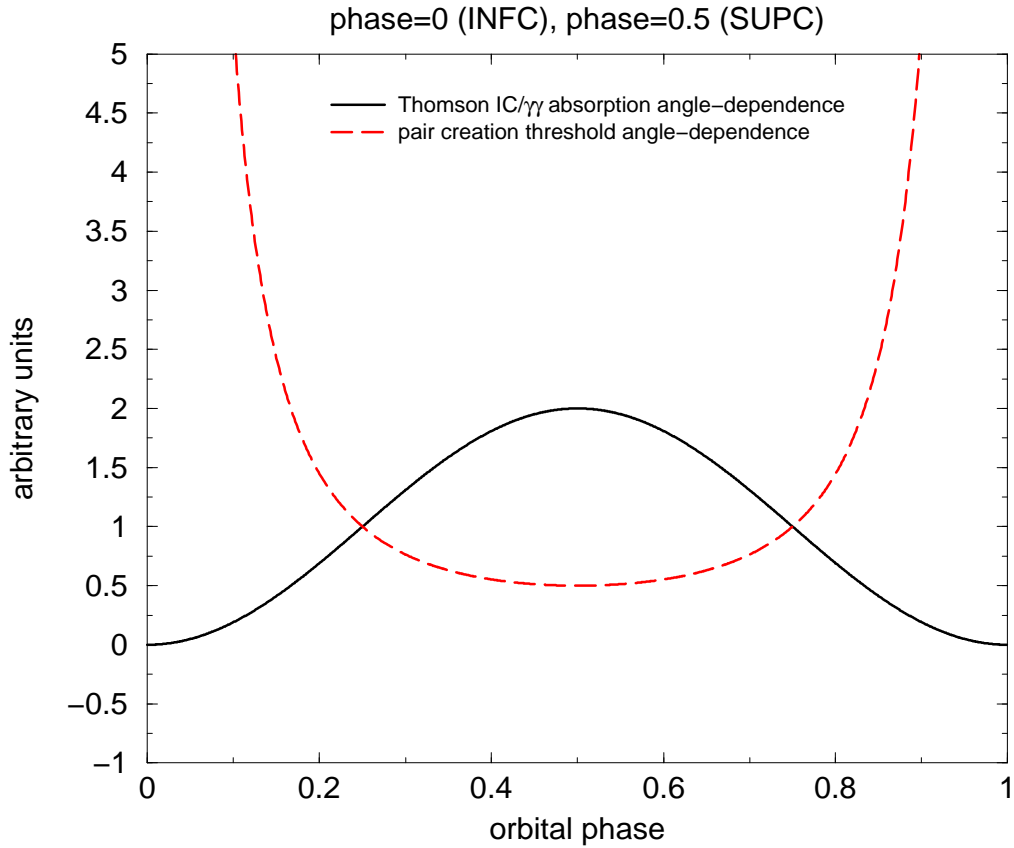,width=13.5cm}
%%          \hspace*{0.3cm}
%   \psfig{figure=fig1b.eps,width=6.3cm}       
\caption{Example of the impact of the geometry variation along the orbit on both photon-photon absorption and IC scattering. 
The pair creation threshold (long-dashed line) 
can go to very large values when target photons come from 
{\it behind} around the inferior conjunction of the compact object (INFC; phase 0.0). This suppresses photon-photon 
absorption below the (increased) pair creation energy threshold.
Depending on the inclination, i.e. the observer line of sight angle with respect to the orbital plane perpendicular direction, 
the stellar target photons can come, 
{\it more or less, from behind}. The case represented here implies an extreme inclination angle of 90$^{\circ}$ for illustrative purposes. 
Also, photon-photon absorption (solid line) also gets smaller when the stellar photons come from {\it behind} (INFC), and gets larger 
when target photons come from the front, around the superior conjunction (SUPC; phase 0.5). The IC scattering rate (also 
solid line), mainly in the Thomson regime, has 
the same dependence on the scattering angle as pair creation, and therefore it gets smaller around INFC and larger around SUPC. \label{lc}}
\end{figure*}

Fig.~\ref{lc} gives an idea of the importance of the angle dependence of photon-photon absorption and IC scattering for the lightcurve. Unlike other sources of variability, this
modulation occurs for sure if the star is providing the targets for these interaction processes.

\section{Final comments}\label{sum}

In this lecture, we try to show that to understand microquasars, first one can approach the problem using simple tools. This can be performed, partially, using the available
phenomenological knowledge on the source properties (known elements of the system/emitter and their relationships, observational information about the emission behavior). This
allows the construction of phenomenological jet emission models which, at some point, fail to deepen in our understanding of microquasar non-thermal radiation. Therefore, in
addition to this {\it phenomenological modeling}, which allows us to set up the basic scenario and obtain some general idea of the global radiation behavior, a simple but
systematic physical study source by source is also to be done. With this study, important constraints can be imposed on the particle acceleration efficiency, the impact of
particle confinement and propagation, and the plausible radiation and attenuation processes occurring in microquasars. 

Once the previous step is finished, it will be time to perform hydrodynamical and magnetohydrodynamical simulations, which are powerful heuristic tools to investigate further the
complexity of the source structure (e.g. to understand the evolution of the jet, and its interaction with the stellar wind and large scale environment, etc.). Finally, using high
quality data observations, provided by the next generation instruments, and the theoretical knowledge acquired with previous modeling and simulations, one can attack the problem
of which mechanism of particle acceleration is more suitable to explain the observed radiation, and the radiation mechanism itself, in the context of a more constraint physical
scenario.

\acknowledgments 
V.B-R. wants to thank the organizers of the school for the opportunity of giving this lecture 
on microquasar physics in such a nice environment.
V.B-R. thanks A.~T. Araudo for her beneficial influence on this work. Finally, V.B-R gratefully acknowledges the 
help of F.~A. Aharonian, 
S.~R. Kelner, D.~Khangulyan, J.~M. Paredes, M. Rib\'o and G.~E. Romero, in the development of the ideas 
presented in this lecture. 
V.B-R. acknowledges also support from the
Alexander von Humboldt Foundation. V.B-R. acknowledges support by DGI of MEC under grant AYA2007-68034-C03-01 and FEDER funds.

\end{document}